\documentclass[aps,prl,reprint,groupedaddress]{revtex4-1}
\usepackage{epsfig}
\usepackage{float}

\begin{document}


\title{EPR's reality criterion and quantum realism}

\author{Florin Moldoveanu}
\affiliation{Committee for Philosophy and the Sciences, University of Maryland, College Park, MD 20742}

\date{\today}

\begin{abstract}
We show that EPR's criterion of reality leads to contradictions in quantum mechanics. When locality is assumed, an inequality involving only one particle is violated, while when parameter and outcome dependence are assumed, EPR-realism is shown to be not Lorentz invariant. Quantum mechanics is both non-local and non-realistic.
\end{abstract}

\pacs{03.65.Ta, 03.65.Ud}


\maketitle


\section{Introduction} 
In 1935 Einstein, Podolsky, and Rosen had introduced their famous argument against the completeness of quantum mechanics \cite{EPR}. Central to their claim was the usage of their definition of a reality criterion stated as follows: {\em ``If, without in any way disturbing a system, we can predict with certainty (i.e., with probability equal to unity) the value of a physical quantity, then there exists an element of physical reality corresponding to this physical quantity''}. 

Following the introduction of the criterion, EPR asserts: {\em `` Regarded not as a necessary, but merely as a sufficient, condition of reality, this criterion is in agreement with classical as well as quantum-mechanical ideas of reality''}. In the end of the EPR paper, the authors state: {\em ``We are thus forced to conclude that the quantum-mechanical description of physical reality given by wave functions is not complete. One could object to this conclusion on the grounds that our criterion of reality is not sufficiently restrictive''}.

The usual explanation of why the EPR paper conclusion is incorrect is based on Bell's analysis\cite{Bell} because the critical assumption of locality prevents obtaining correlations predicted by quantum mechanics. However, it is the aim of this paper to show that realism as described by the EPR reality criterion is not universally valid as well and therefore both locality and realism are violated by nature. 

A popular interpretation of Bell's theorem is that local realism is false. Care must be exercised when speaking of realism or locality because there are distinct levels of them. For example, on the non-locality side there is Bell-locality, non-signaling, quantum field theory micro-causality, or the Bohm-Aharonov effect \cite{Aharonov1}. For the realism side, there are various philosophical distinctions of realism, but the EPR-reality criterion has probably the sharpest definition which can withstand mathematical arguments.   

Different interpretations of quantum mechanics can exhibit a tradeoff between locality, realism, and counterfactual definiteness. For example, the de Broglie-Bohm interpretation is considered realistic, but non-local \cite{Goldstein1}, and the many-worlds interpretation is both realistic and  (mostly) local \cite{Vaidman1}, but violates counterfactual definiteness. 

In light of those results, it seems misguided that realism is at fault as well in quantum mechanics, particularly due to de Broglie-Bohm interpretation. However, in this interpretation spin is treated just like in standard quantum mechanics with the justification that spin has no classical counterpart \cite{Goldstein1}. This paper will show that for spin, the EPR reality criterion leads to contradictions and hence this criterion is not universally valid. 

\section{EPR reality criterion contradictions}
When measuring a physical parameter, by the collapse postulate we can predict with certainly that a repeated measurement done in quick succession yields the same value. Thus one can assert that the physical system does indeed have that particular physical parameter value. The reverse implication is given by the EPR reality criterion: if we can predict with certainly the outcome of an experiment, there must be an element of reality associated with that outcome. 

To prove the EPR reality criterion generates contradictions in the spin case, we will start arguing along the lines of Bell. Let us start with one electron and proceed to perform the standard spin measurement experiment using the Stern Gerlach device. Now without disturbing the system, we can predict with certainty that we will get either a positive or negative deflection, and thus by EPR reality criterion, there must exist an element of physical reality corresponding to this physical quantity. For lack of a better word, let us call this element of physical reality ``spin''. What are its characteristics? Inheriting its characteristics from the measurement process, spin must have a direction and a binary outcome. Those may not necessarily correspond to intrinsic properties, but if we assume counterfactual definiteness, when we say spin has a definite direction and a definite value, this is actually shorthand for the statement that if we were to position a Stern Gerlach apparatus on that specific direction we would obtain with certainty that particular deflection value. 

What else do we know? If we perform two subsequent measurements, first on a direction $m$ and second on a direction $n$, the probability that the second measurement obtains the same outcome as the first measurement is cosine square of the half angle between the two directions: $\cos^2 (\alpha_{mn} /2)$. This is an additional physical property which will allow us to construct arguments disproving the EPR reality criterion. 

One way we can understand Bell's result is that existence can be tested by two methods: direct measurement or correlation analysis. In the quantum mechanics case, due to superposition, the two methods do not necessarily agree. To prove the rejection of realism (and not the weaker version of local realism), we need to pursue single particle arguments. However, since hidden variables models for a single electron do exist, we will pursue single particle arguments when superposition is also present. What we would like to do is obtain a Bell-type inequality concerning a single particle.

Let us consider at this point the usual singlet Bell state: $\Psi = 1/\sqrt{2}(|+\rangle |-\rangle-|-\rangle|+\rangle)$ in the EPR-B setting. When measuring the spins on two directions $a$ and $b$, the correlation for this singlet state is $-a \cdot b$. In particular, when the two directions are completely opposite, one gets perfect correlations between the two measurements and this is a very demanding constraint to be obeyed by any hidden variable model.

Suppose we measure the spin values for the two particles in the EPR-B singlet state on the vertical axis. Without disturbing the system, we know with certainty before performing the measurement that that we will either obtain $|+\rangle |-\rangle $ or $|-\rangle |+\rangle $. By EPR reality criterion, the spins should be either $|+\rangle |-\rangle $ or $|-\rangle |+\rangle $. When locality is assumed, we will show that neither of the two possibilities is allowed by the derivation of an inequality. A different problem will arise when locality is not obeyed.

Suppose that in the original singlet state the spins take the definite values of $|+\rangle |-\rangle $ (the other case is treated identically). Instead of measuring on the vertical axis, let us measure the two spins on two orthogonal directions each making 45 degrees with the vertical axis. 

The correlation between the two new measurements is zero because the measurement directions are orthogonal. Can a hidden variable model obeying the consecutive measurement law of $\cos^2 (\alpha_{mn} /2)$ generate a zero correlation when the initial state is either $|+\rangle |-\rangle $ or $|-\rangle |+\rangle $? Suppose we repeat the experiment $N$ times on a system prepared as $|+\rangle |-\rangle $ and we group together the identical and different outcomes. Suppose Alice obtains $P$ positive outcomes and $N$ negative outcomes (and the same holds true for Bob). Also suppose for $Q$ outcomes Alice obtains positive measurements and Bob obtains negative measurements. The correlation is $+(P-Q) - Q + (N-Q) -Q = P+N-4Q$. In order to obtain zero correlation, the $Q$ outcome should occur 25 percent of time but this is impossible because at the same time the following inequalities must hold: $Q<P$ and $Q<N$. The same outcomes are obtained $ cos^2 (\pi /8) $ percent of time, and the different outcomes are obtained $ sin^2 (\pi /8) $ percent of time. However, $1/4 > \sin^2 (\pi /8)\approx 0.1464$ and hence we have a contradiction. This result holds for both $|+\rangle |-\rangle $ and $|-\rangle |+\rangle $ cases.

\begin{figure}[h]
\centering
\epsfig{file=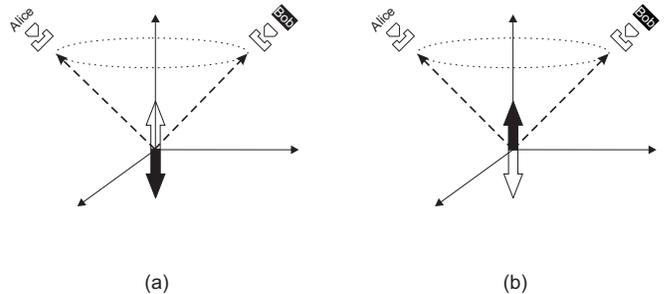,width=1\linewidth,clip=}
\caption{Spin alignment: (a) Alice's electron spin point up and Bob's electron spin points down. (b) Alice's electron spin point down and Bob's electron spin points up.}
\label{fig:45Degrees}
\end{figure}

One obvious counterargument is that we had assumed the spins to be aligned on the vertical axis because if we would measure on this particular axis we would get an outcome on this axis. From the point of view of an intrinsic property this looks conspiratorial. However strange, the assumption of the spin to have a definite value along the vertical axis is a consequence of our definition of spin as a reflection of the measurement process together with the EPR reality criterion. EPR reality criterion and locality demands that we should not see any difference measuring on two orthogonal directions making $\pi /4$ angle with the vertical axis irrespective of measuring or not first on the vertical axis. The key point is that while we cannot predict with certainty the outcome of the intermediate result when measuring on the vertical axis, the inequality holds for {\em all} possible outcomes of the intermediate measurement and we can apply the EPR reality criterion to all of them arriving at the contradiction. 

It is easy to explain why there is a contradiction: measuring first on the vertical axis collapses the state and the non-commutativity of spin measurements yields different outcomes. Still, there are two questions. First, can we obtain this inequality (which is about individual outcomes and not correlations) with only one particle? Second, what makes this inequality possible? The answer to the first question is negative but the mathematical argument is rather long and unsurprising given the existence of hidden variable models and will not be presented here. The answer to the second question is more interesting because the root cause is quantum superposition. Comparing the intermediate states demanded by the EPR reality criterion with the singlet state, the difference is in quantum superposition, or in the difference between classical and quantum disjunction. Because quantum mechanics obeys the logic of projective spaces and not that of set theory, quantum disjunction has non-classical characteristics \cite{Aerts1}. In particular two statements $A=|+\rangle |-\rangle $ and $B=|-\rangle |+\rangle $, can be both false and yet $A \vee B_{|\phi} = 1/\sqrt{2}(|+\rangle |-\rangle + e^{i \phi} |-\rangle|+\rangle) $ can still be true. 

Bell explained that any local realistic model for the singlet state develops a ``kink'' in the correlation of probabilities for aligned measurement directions, while the slope of the correlation as a function of angle in the quantum mechanics case is zero due to superposition. The same effect happens here and there will be no inequality contradiction for a single particle if superposition was not present. 

Since we had assumed locality so far, at this point the proof that EPR reality criterion leads to contradictions is incomplete. We could construct non-local explanations which as an added property will no longer demand the spin to be aligned on the vertical axis in the beginning.   

Any electron spin hidden variable model should obey the law of consecutive measurements: $\cos^2 (\alpha_{mn} /2)$ and this depends only on the angle between the two consecutive measurement directions. But this angle may depend in turn on other parameters. In general, Bell locality (or separability) is equivalent with parameter and outcome independence. In the EPR-B experiment, suppose that spins have a definite value before measurement (counterfactual definiteness). Also suppose the two spins are not aligned on the same axis. If we happen to measure the spin for Alice particle on a direction aligned with her intrinsic particle spin direction, and later we attempt to measure Bob's particle on the opposite direction, the outcomes must be perfectly correlated. However, for Bob the outcome will not be perfectly correlated because he will agree with Alice's value only cosine square of the half angle between the two intrinsic spin directions. Moreover, no outcome or parameter dependence could help Bob achieve perfect correlation because Alice just happens to measure on the same direction as her particle intrinsic spin direction and she will not disturb her spin configuration in any way (and this in turn cannot change Bob's spin orientation). What this argument shows is that if the electrons have definite spins before measurement, in order to obey both the law of consecutive measurements and Bell correlations, in the singlet state the two spins must be aligned on the same direction, either pointing away from each other, or towards each other.

At this point we could consider with Bell an isotropic distribution of opposite spins and compute the overall correlation arriving at $-1/3 (a \cdot b)$ which contradicts the quantum mechanics prediction. But this would only prove that Bell locality is violated and hence it will not be a new result.     

Having established this property of intrinsic spin direction, we should first attempt to eliminate the contrived need for the two spins to be aligned on the vertical axis. Indeed, if Alice measures first (and say she obtains $+1$), and if this measurement changes instantaneously the direction for the spin of the electron Bob will measure next aligning it with Alice's measurement direction, then the law of subsequent measurements yields Bell correlation: Bob will get $-1$ $cos^2 (\alpha_{ab} /2)$ percent of the time and $+1$ $sin^2 (\alpha_{ab} /2)$ percent of the time for the correct correlation: $- a \cdot b$ .

Now we need to show that this model of obtaining Bell's correlation is unique. Indeed, the correlation between subsequent measurements can only depend on the angle between the measurement directions. Parameter and outcome independence results in the earlier inequality contradiction. Parameter dependence can change the orientation of the local spin based on the remote orientation of the other measuring device. Outcome dependence is irrelevant because of the periodicity of the sine and cosine functions which yields the same conclusions regardless of the intrinsic spin direction: up or down on a measurement orientation. The only trigonometric identity able to obtain the singlet state correlation based on the subsequent measurement rule is the one presented above. 

But in the non-local case there are problems defining existence in a unique way (it is not Lorentz invariant). The root cause is the ambiguity on which measurement changes the remote particle. If Alice does the measurement first, this will affect Bob's electron spin orientation, and the other way around. But if Alice and Bob are spacelike separated, in a reference frame Alice performs the measurement first, and in another Bob does it first. If the angles of the two measurements are arbitrary, the evolution of say Alice's electron spin orientation is not observer independent and in a reference frame the spin will have a definite value along a direction, while in another reference frame the spin will never have had that value. As such, existence is reference frame dependent contradicting EPR reality criterion. 
\begin{figure}[h]
\centering
\epsfig{file=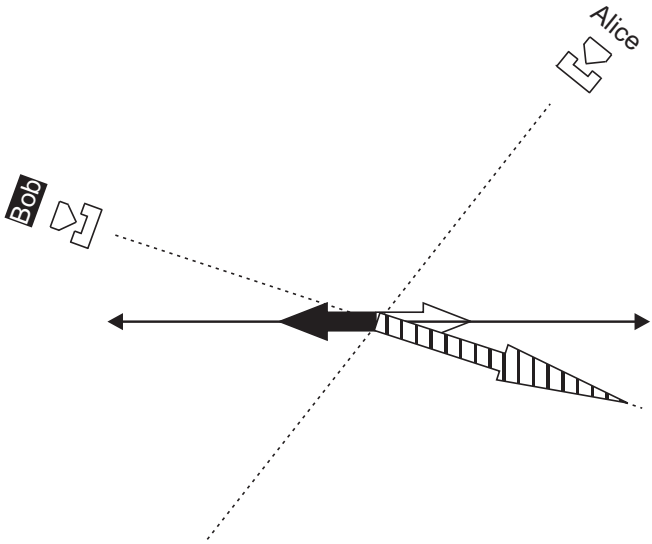,width=1\linewidth,clip=}
\caption{Originally the spins are aligned on the horizontal line. If Bob is first to perform the measurement this will align the electron spin for Alice on Bob's measurement direction (the pattern-filled arrow). But if Alice measures first, the spin of her electron would never have had that orientation.}\label{fig:Lorentz}
\end{figure}

One may object that introducing Lorentz invariance in the discussion is inappropriate in the context of non-relativistic quantum mechanics. However, different correlations models were proposed in the past in a similar context, the so-called ``before-before'' model \cite{Suarez1} and experimental result \cite{BB_experiment} ruled it out in favor of non-relativistic quantum mechanics results, validating the applicability of the argument above.

\section{Conclusion}

To summarize, EPR reality criterion demands the objective existence of spin independent of measurement. Parameter and outcome independence yields a contradiction based on an inequality. Parameter and outcome dependence can explain the singlet state correlation but contradicts Lorentz invariance (this argument is similar with Hardy's argument \cite{HardyParadox}). For the non-local and realistic Bohmian interpretation, the present result does not constitute a disproof, but reduces the appeal of this approach, because paying the price of ``surreal'' trajectories \cite{surreal1} does not buy complete objective reality. 

Different interpretations of quantum mechanics can succeed in providing realism or locality to various degrees, but never complete. In the many-worlds interpretation, locality is mostly obeyed, and Bell inequalities are irrelevant as occurring in different branches. 

In quantum field theory, based on micro-causality, a possible interpretation is that quantum mechanics is local, but not realistic. However, this is incorrect, because micro-causality is a relativistic quantum field theory axiom independent of standard non-relativistic quantum mechanics.  

A recently introduced argument \cite{PBRPaper} proves that quantum mechanics is not epistemological. This argument shows that quantum mechanics is not purely ontological either.

It is safe to state that quantum mechanics is both non-local and non-realistic with different levels of non-locality and non-realism based on interpretation. 

\section{Acknowledgement} I want to thank Marc Holman for helpful correspondence. He originally introduced the idea of consecutive measurements to refute a challenge to Bell's theorem, and this led to a systematic investigation of its consequences.

\end{document}